# qqH Search at the LHC
# With H Decay to Two Leptons and Two Neutrinos

Dan Green

Fermilab

February, 2005

**Abstract**

The search for the Higgs boson produced by Vector Boson Fusion (VBF) where the Higgs decays into W pairs and thence to leptons plus neutrinos is discussed at the Monte Carlo generator level. In particular, the cuts which are applied in order to reduce backgrounds are almost exactly those used to extract the VBF production of Z bosons.

**Introduction**

There have been many studies of the VBF production of Higgs bosons. [1]. In particular, there were preliminary studies using the CMS Monte Carlo with fairly complete detector simulation which were performed some time ago [2]. These studies confirmed that the VBF process is a very promising reaction for Higgs searches. Presently this process is planned to be explored in greater depth for the upcoming CMS Physics Technical Design Report. In order to attempt to first isolate the VBF process, the production of (qqZ) via the VBF mechanism has been studied at the generator level [3] and a series of background cuts were developed.

The (qqZ) and (qqH) processes for VBF have the same Feynman diagrams because of the similar WWZ and WWH couplings. If leptonic decays are studied for the Z and for the WW decays of the H, then both processes will appear in the di-lepton plus jets data stream. Therefore, the relative cross sections will have reduced systematic uncertainties due to trigger biases. In addition, strong trigger cuts should not be necessary on the "tag jets", [4] so that their characteristics can be studied in an offline analysis using the (qqZ) sample.

Because the VBF process has not yet been definitively isolated in Z plus jets production, it is important to first establish that this (qqZ) process can be isolated at the LHC. Because the effective (qqZ) cross section after cuts is larger than that for (qqH), the "standard candle" for VBF can be established prior to searching for the (qqH) signal.

**VBF Production and Backgrounds**

The VBF production of Higgs at the LHC is at least 1/10 of the full cross section for Higgs production [5]. The remnant outgoing jets at small angles to the beam, the "tag jets", allow for a large increase in the signal to background ratio with respect to inclusive Higgs production. If the WW decays are measured in the final state, then the WWH coupling is isolated and can be determined. In this work the decays of W to lepton plus neutrino occur isotropically which must later be modified to reflect the proper (V-A) weak decay dynamics.

The measurement of the tag jets also gives additional information. For example, the final state total tag jet longitudinal momentum gives an approximate measure of the initial state longitudinal momentum because the radiated W are low momentum, as seen in Fig.1. The tag jets also determine the Higgs transverse momentum, since the initial state can be assumed to have small transverse momentum.

The final state studied here is W pairs plus two small angle, or "tag" jets. Clearly, one large background will come from strong top pair production where the two b jets mimic the tag jets. This process is called (tt) in what follows. Another possible background is the production of a top pair accompanied by a radiated gluon, called (gtt) here. This gluon is typically at small angle with respect to the beam, and more readily mimics a "tag" jet. The extra b jet must then be at a low enough transverse momentum to escape being detected as a jet and thence rejected. The (gtt) process was modeled using the COMPHEP [6] program, where the evaluated Feynman diagrams are shown in Fig.2. There is also a background process initiated by gluon plus valence quark which leads to a small angle quark in the final state, called (qtt) in what follows. Although the uncut cross section is smaller than that for (gtt), the valence quark more effectively mimics a tag jet and the cuts are less effective in removing this background.

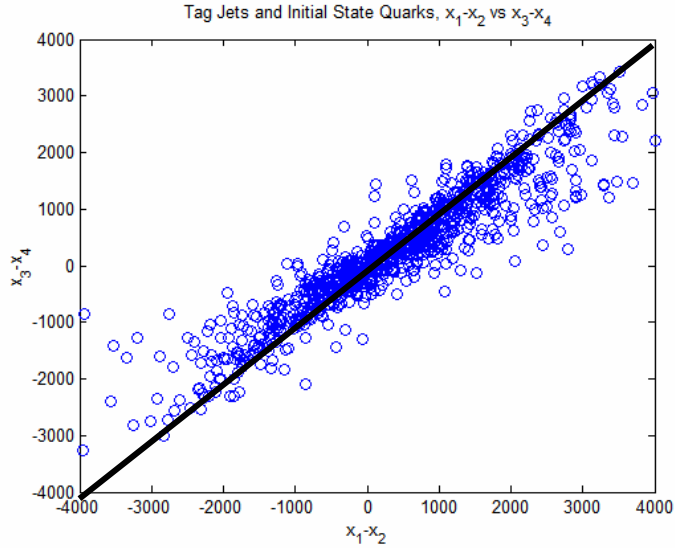

Figure 1: Relationship between the initial state longitudinal momentum and that of the final state in VBF production of a light Higgs boson. The line assumes initial-final state equality.

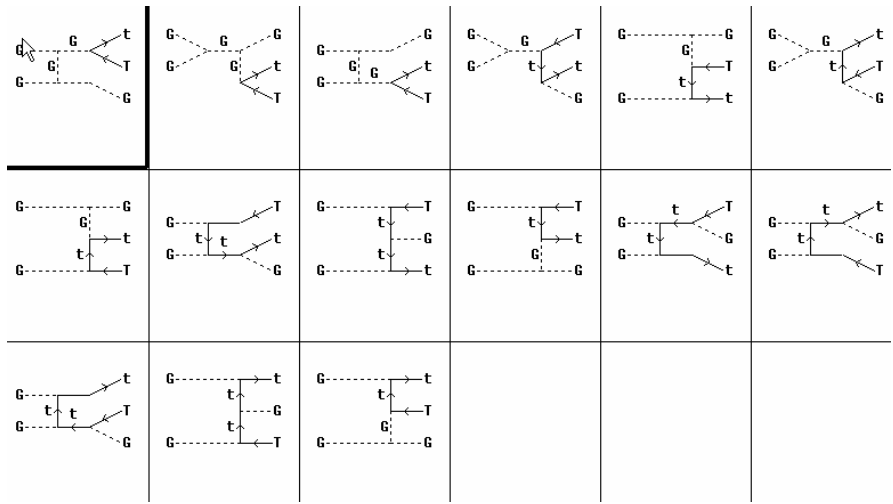

Figure 2: Feynman diagrams for the production of a top quark pair plus a gluon, (gtt).

Other backgrounds have yet smaller uncut cross sections. They consist of electroweak production of W pairs with additional jets. The cross section for W pairs has recently been reported by CDF [7]. The COMPHEP evaluation of the Drell-Yan production of W pairs agrees fairly well with the measured cross section. Extrapolating in energy to the LHC and adding two radiated gluons leads to a cross section almost one hundred times less than the (gtt) cross section. However, there are processes where valence quarks electroweakly produce W pairs plus jets. In the case of two valence quarks, there is a continuum background of VBF produced W pairs, which is a small background to the resonant Higgs production. There is also the case of one valence quark and one gluon in the initial state, called (qgWW) in what follows. This process appears to be the largest electroweak background as evaluated by COMPHEP and the higher x valence quarks make this process more resistant to the applied cuts.

**Cuts and Background Reduction**

In order to reduce systematic errors on the ratio of (qqZ) and (qqH) production rates, the same cuts are applied in the two cases. The resulting cross sections for both signal and background are shown in Table 1 below.

**Table 1**
**Cut Study for VBF Produced Higgs and Backgrounds**
**[Cross Sections in pb, Mass in GeV]**

| cut | gtt | tt | qtt | qgWW | qqH |
|---|---|---|---|---|---|
| none | 3700 | 1320 | 460 | 120 | 3 |
| $y_1=(-5,-1)*y_2=(1,5)$ | 490 | 215 | 56 | 15 | 2 |
| $<M_{12}>$ | 316 | 320 | 650 | 340 | 1330 |
| $M_{12} > 750$ | 28 | 2.1 | 17 | 1.1 | 1.4 |
| $<P_{tb2}>$ | 98 | | 92 | | |
| $P_{tb2} < 20$ | 1.2 | 2.1 | 0.62 | 1.1 | 1.4 |
| $(y_2-y_H)>1.5*(y_H-y_1)>1.5$ | 0.33 | 0.38 | 0.14 | 0.18 | 0.84 |
| $M_{1l}>160*M_{2l}>160$ | <0.07 | <0.013 | <0.005 | <0.08 | 0.34 |
| $P_{TH}$ | 115 | 75 | | | 130 |
| $M_H$ | 406 | 490 | 535 | 360 | 170 |

In Table 1 the processes are ordered by decreasing uncut cross section. The three largest are due to strong production of top pairs, followed by electroweak production of W pairs by a valence quark plus a gluon. Without cuts the (qqH) signal is buried by a factor about 1000. The cuts used are those applied in the (qqZ) case. First there are cuts on the pseudorapidity, called y here, of the tag jets. There should be one in the forward hemisphere and one in the backward. The scatter plot for the signal and the largest background processes are shown in Fig.3. This cut improves the signal to background ration to be about 1/250, or by a factor of about four.

The next sequential cut is on the mass parameter for the two tag jets. It is defined in Eq. 1 to be the "parallel" mass where the mass contribution of the transverse momentum is retained because the tag jets from (qqH) typically have larger transverse momentum than the background processes. The mean value of the tag pair mass after the first cut is given in Table 1. This cut follows from a prior study of the appropriate mass variables [8]

$$M_{12} = (E_1 + E_2)^2 - (P_{\|1} + P_{\|2})^2 \qquad (1)$$

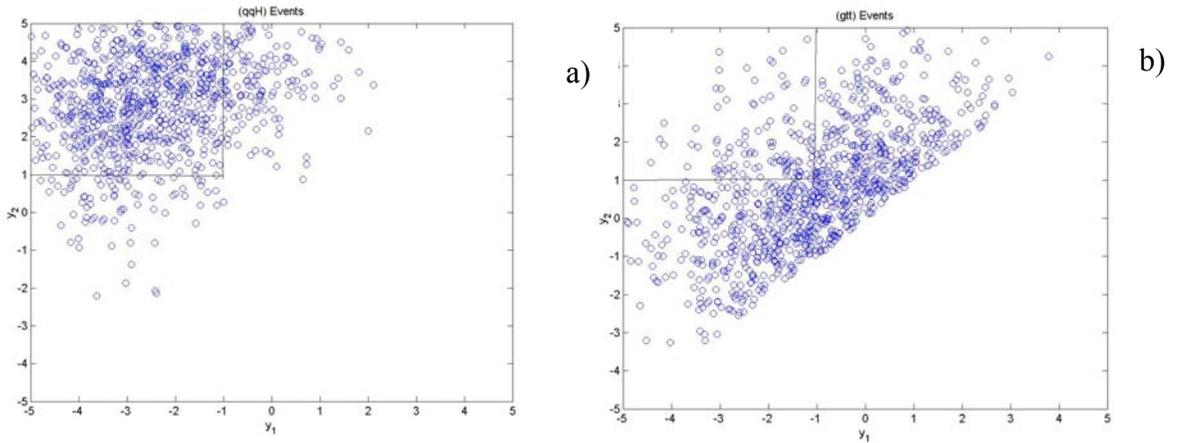

Figure 3: Scatter plot of the rapidity of the two tag jets in a) - (qqH) and b) - (gtt) events. The lines indicate the imposed cuts. The tag jet labeled as 1 is the trailing one while tag jet 2 is leading, i.e. $y_2 > y_1$.

The distribution of tag jet masses is shown in Fig. 4. Note that the mean of the signal tag mass is roughly 4 times the background except for (qtt) where it is only about twice as large.

After the tag mass of the pair is required to be greater than 750 GeV, the signal to background ratio is about 1/20 which is an improvement by a multiplicative factor of twelve. The two largest remaining backgrounds, (gtt) and (qtt) contain an extra jet. This jet has an average transverse momentum of about 95 GeV. If it is assumed that a 20 GeV jet can be resolved from the pileup energy due to the underlying event and pileup from other inelastic events within the bunch crossing, then a "jet veto" of 20 GeV can be made. It is also assumed that signal events do not suffer losses in efficiency due to the presence of false jets, for example from initial/final state radiation (ISR/FSR). The transverse momentum distribution of the extra b jet is shown in Fig.5. Note that this particular cut is specific to the backgrounds considered here and is not part of the (qqZ) sequential cuts.

The signal to sum of backgrounds is about 3.6 at this point, indicating an improvement by another factor of about twenty five for (gtt) and (qtt) events.

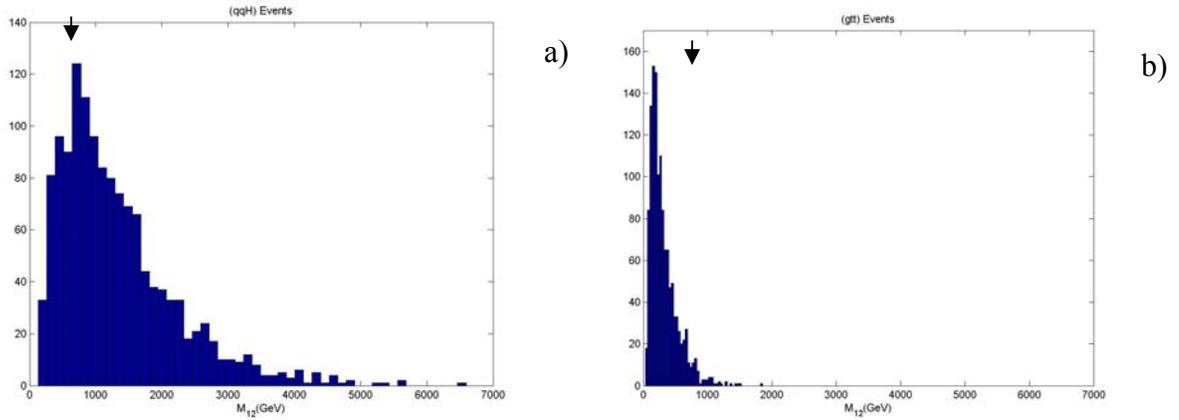

Figure 4: Mass parameter for the two tag jets for a) – (qqH) and b) – (gtt) events. The arrows indicate the cut value of 750 GeV.

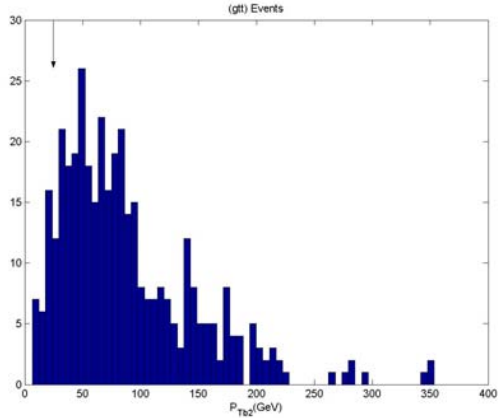

Figure 5: Transverse momentum of the additional b quark in (gtt) events. The arrow indicates the cut value.

Following the (qqZ) strategy, the next cut to be imposed requires "angular order" of the two tag jets and the "Higgs". The "Higgs" is defined by the tag jet and decay di-lepton kinematics, along with the missing transverse energy due to the final state neutrinos. In the (qqH) process, there is a correlation between the di-lepton longitudinal momentum and the Higgs longitudinal momentum, as shown in Fig.6 for COMPHEP generated events.

Using the approximation shown in Fig.6, that the Higgs longitudinal momentum is twice that of the di-lepton system, the four momentum vectors of the di-neutrino, $P_{\nu\nu}$ and di-lepton, $P_{\ell\ell}$, systems can be approximated as shown in Eq.2.

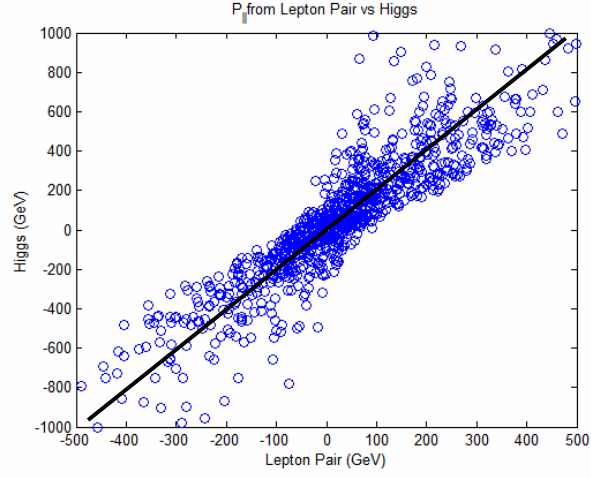

Figure 6: Correlation between the di-lepton and Higgs longitudinal momentum for a 180 GeV Higgs using the COMPHEP program. The line indicates the approximation that the Higgs has twice the di-lepton longitudinal momentum.

$$P_{\nu\nu} = (E_T, P_{\ell 1z} + P_{\ell 2z}, m_{\ell 1 \ell 2})$$
$$P_{\ell\ell} = (\vec{P}_{T\ell 1} + \vec{P}_{T\ell 2}, P_{\ell 1z} + P_{\ell 2z}, m_{\ell 1 \ell 2})$$
(2)

Taking the "Higgs" to be the sum of the di-neutrino and di-lepton systems, with masses taken from threshold kinematics, [9], $P_H = P_{\nu\nu} + P_{\ell\ell}$, the "Higgs" pseudorapidity can be calculated and used in the angular ordering cut. The resulting cuts regions for both signal and background are shown in Fig. 7.

After the angular ordering cut the signal to background ratio is about 1/1.2. There are additional cuts which can be used to further improve this ratio. One of them is specific to the existence of top background in the (qqH) study as opposed to the (qqZ) backgrounds. Specifically, if the tag jet candidate is in truth a b quark from the top decay, then there is a kinematic maximum value that the mass of the tag jet and the lepton from the cascade W decay can attain. A scatter plot of the leading tag-leading lepton mass versus the trailing tag – trailing lepton mass is shown in Fig. 8 for signal and background, (tt), events. The kinematic edge is very evident for the (tt) background.

The distribution for (qqH) and (gtt) events is shown in Fig.9. In this case only one lepton-tag combination is associated with a top decay. The other tag jet is the gluon. Clearly the signal to background ratio can be greatly improved for backgrounds where the W pairs arise from top quark decays. In fact, as shown in Table 1, there are no remaining events after this sequential cut is imposed and the signal to sum of backgrounds is greater than 2.3.

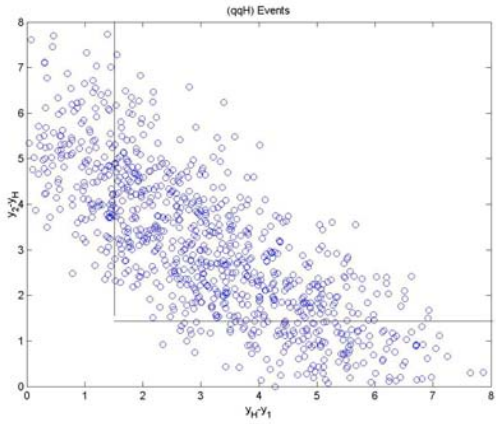
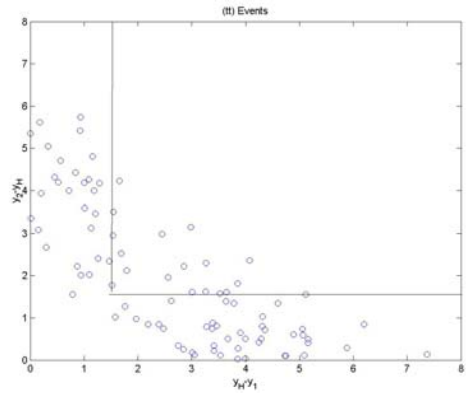
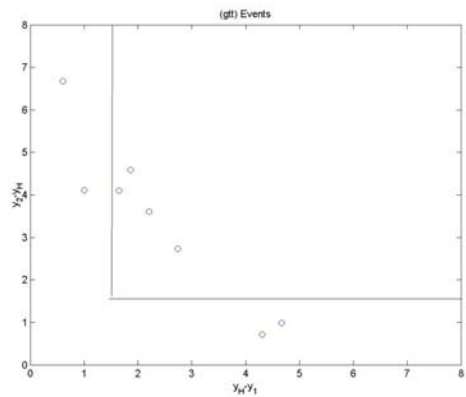

Figure 7: Scatter plot of the pseudorapidity difference between the leading tag jet and the "Higgs" and the pseudorapidity difference between the trailing lepton and the "Higgs" for a) – (qqH) events, b) – (tt) events, and c) – (gtt) events.

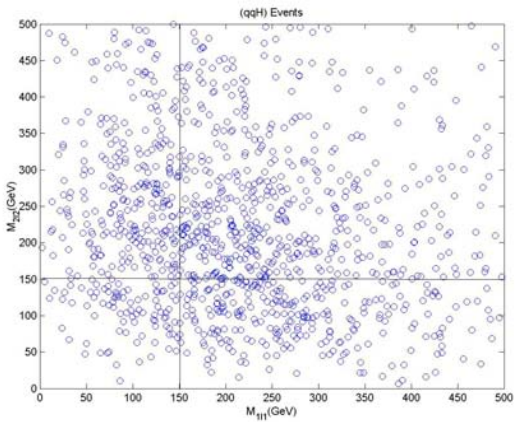
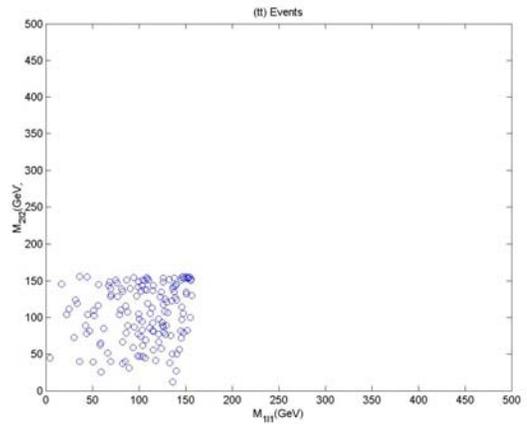

Figure 8: Scatter plot of the mass of the leading tag jet and the leading lepton and the trailing tag jet and the trailing lepton for a) – (qqH) events, and b) – (tt) events.

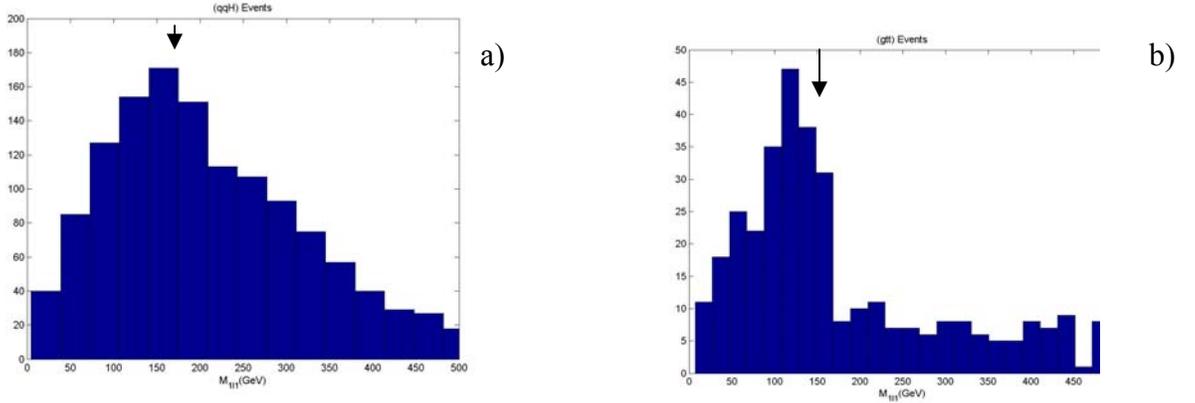

Figure 9: Mass of the leading lepton and the leading tag jet for a) –(qqH) and b) - (gtt) events. The arrows indicate the cut value imposed.

In the study of (qqZ) VBF a cut on the transverse momentum of the Z was used to improve the background rejection. For the (qqH) case, the distribution for signal and background is shown in Fig.10 where the cut after angular ordering is not imposed in order to retain some events. Clearly, there can be improved background rejection by imposing a "Higgs" transverse momentum cut. The mean value of the "Higgs" transverse momentum is given in Table 1.

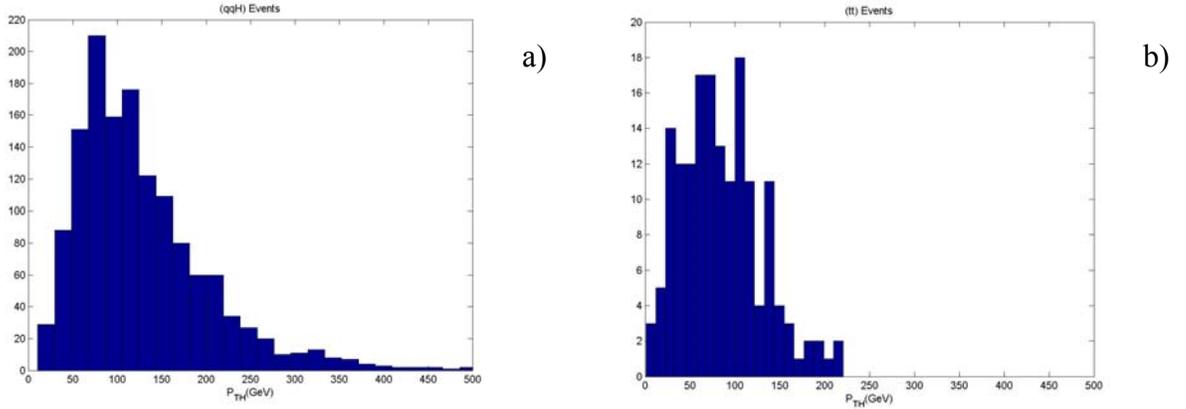

Figure 10: Transverse momentum of the "Higgs" for a) - (qqH) and b – (tt) events.

Finally, the "Higgs" mass can be used in looking for a resonant signal. The mass of the "Higgs", defined in Eq. 3, is a slight extension of the successful transverse mass variable used previously [9] in order to include an estimate for the di-neutrino longitudinal momentum.

$$M_H^2 = (E_{\nu\nu} + E_{\ell\ell})^2 - (\vec{P}_{\nu\nu} + \vec{P}_{\ell\ell})^2 \qquad (3)$$

The distribution of the "Higgs" mass parameter is shown in Fig.11. Clearly, there is a resonant peak for the signal, which was generated at a mass of 180 GeV in this study. The r.m.s. of the mass distribution is about 45 GeV. The mean value for the background mass parameters is given in Table 1 and the distribution with relaxed cuts is shown in Fig. 11. It is clear that if there are sufficient events and if the imposed cuts yield a signal to background about one, then the Higgs resonance will be evident in the data. Note that in Table 1 the branching fraction for H -> WW has been assumed to be equal to one. Lower values can easily be assumed in evaluating the cuts.

For the case of electroweakly produced W pairs, the cross sections are smaller than the main backgrounds by an order of magnitude. However, after cuts on tag jet pseudorapidities, tag pair masses and extra jet vetoing, this background is comparable to that for the (gtt) and (tt) processes. The detailed effect of harder cuts requires that a larger background event sample be generated.

**Conclusions**

The (qqZ) data sample will allow the LHC experiments to define a "standard candle" for isolating the VBF process. There are serious backgrounds, but they can be overcome with the imposition of a few cuts to attain a signal to background ratio near one. These same cuts can then be imposed on (qqH) and background events. Additional cuts arise from the fact that the (qqH) backgrounds largely are due to events containing top pair decays. In turn that implies the veto of extra jets in the event and the limitation of the lepton-tag mass in top background events.

The (qqZ) cross section before/after cuts which yield a roughly one to one signal to background goes from 14.4 (uncut) to 2.4 pb. In the (qqH) case the reduction in signal for a similar signal to background ratio is from 3.0 pb (uncut) to 0.84 pb. Assuming a Higgs branching ratio into W pairs of 1.0 and allowing only direct electron or muon decay of the Z and the two W's, the (qqZ) di-lepton cross section is 0.16 pb, while the (qqH) signal is 0.038 pb. Therefore, the (qqZ) events will be seen early on in a clean resonant Z sample with two additional jets. This sample can be used to study the tag jet cuts which can then be applied to the (qqH) candidates which would appear in the di-lepton plus two jet plus missing transverse energy sample.

**References**

1. K. Iordanidis and D. Zeppenfeld, hep-ph/9709506,
   D. Rainwater and D. Zeppenfeld, hep-ph/9906218
   K. Cranmer et al., ATL-PHYS-2003-002, Jan., 2003
   S. Asai et al., hep-ph/0402254, Feb., 2004
   B. Mellado, hep-ex/0405043, May 2004
   K, Cranmer et al., ATL-PHYS-2004-019, July, 2004
   E. Berger and J. Campbell, Phys.Rev.D, 70, 073011, Oct., 2004
2. N. Akchurin et al., Fermilab-FN-0714, Feb., 2002

N. Akchurin et al., Fermilab FN-0723, Aug., 2002
3. D. Green, hep-ex/0502009
4. CMS – DAQ Technical Design Report – CERN/LHCC 2002-26, Dec., 2002
5. M Spira and P.M. Zerwas, Lecture Notes in Physics – 512, Berlin, Springer-Verlag (1997)
6. COMPHEP. A. Pukhov et al., Preprint INP MSU 98-41/541 and hep-ph/9908288
7. CDF/DOC/ELECTROWEAK/PUBLIC/7175, Aug., 2004
8. D. Green, hep-ex/0501027
9. N. Kauer, T. Plehn, D. Rainwater, and D. Zeppenfeld, hep-ph/0012351, Dec., 2000

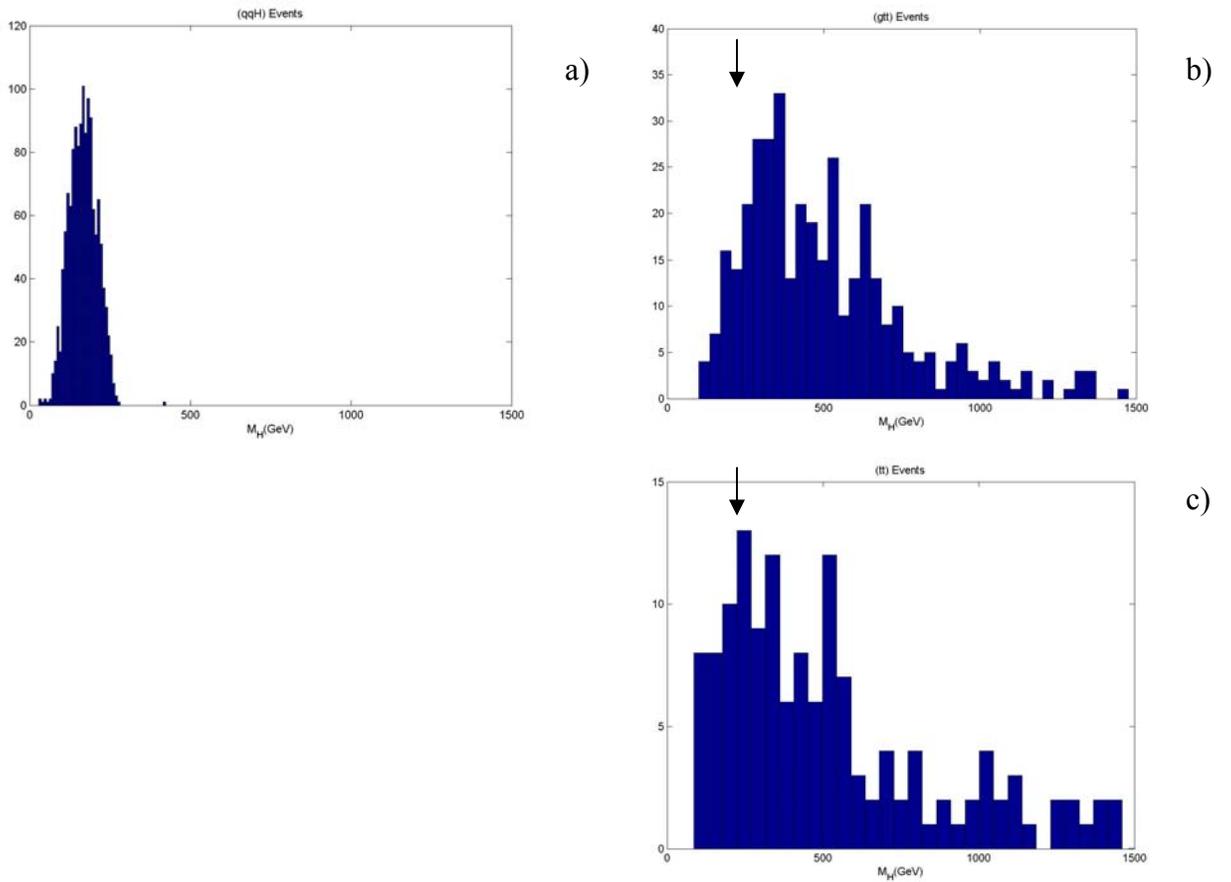

Figure 11: Mass distribution of the "Higgs" for a) – (qqH) events, b) – (gtt) events, and c) – (tt) events. The arrows indicate the peak of the (qqH) distribution.